\documentclass[twocolumn,aps,prl,psfig,showpacs]{revtex4}
\usepackage{amsfonts}
\usepackage{amsmath}
\usepackage{graphicx}
\usepackage{bm}
\usepackage{amssymb}
\usepackage{times}
\usepackage{dcolumn}

\makeatletter

\begin{document}

\title{Family of boron fullerenes: general constructing schemes, electron
counting rule and \emph{ab initio} calculations}
\author{Qing-Bo Yan, Xian-Lei Sheng, Qing-Rong Zheng, Li-Zhi Zhang, Gang Su$%
^{\ast}$}
\affiliation{College of Physical Sciences, Graduate University of Chinese Academy of
Sciences, P. O. Box 4588, Beijing 100049, China}

\begin{abstract}
A set of general constructing schemes are unveiled to predict a large family
of stable boron monoelemental, hollow fullerenes with magic numbers $32+8k$ (%
$k\geq 0$). The remarkable stabilities of these new boron fullerenes
are then studied by intense \emph{ab initio} calculations. An
electron counting rule as well as an isolated hollow rule are
proposed to readily show the high stability and the electronic
bonding property, which are also revealed applicable to a number of
newly predicted boron sheets and nanotubes.
\end{abstract}

\maketitle

\bigskip

Carbon nanostructures such as fullerenes, nanotubes and graphenes \cite%
{c60discover, fullerene_book, nanotube, graphene} have been extensively
studied in the past decades. As boron and carbon are neighbors in the
periodic table, and possess many structural analogies \cite{BCsimilarity1,
BCsimilarity2}, it is expected that boron could also form nanostructures
similar to carbon. The boron clusters, nanotubes (BNT) and sheets are thus
actively explored in recent years \cite{Boustani, Evans, Kunstmann}. More
currently, a stable boron fullerene B$_{80}$ was predicted \cite{B80predict}%
, and a new type of boron sheet (NBS) and related BNTs
\cite{newboronsheet, newBNT} were found to be remarkably more stable
than the boron sheet and BNTs with triangular structures; several
other fullerene-like boron nanostructures are also predicted
\cite{B180, B110}. In this Letter, we shall present generic
constructing schemes that can
produce a large family of novel stable boron nanostructures that include B$%
_{80}$ buckyball and the NBS as members.

Let us start with analyzing the geometrical structures of B$_{80}$ and the
NBS. They contain three basic motifs: hollow pentagon (HP), hollow hexagon
(HH), and filled hexagon (FH) (i.e. hexagon with an additional atom in
center), respectively. A closer inspection reveals that both B$_{80}$ and
NBS are entirely composed of such a snowdrop-like structure [Fig. 1(a)] in
that a central FH is surrounded by three FHs and three "hollows" (HPs or
HHs), and every FH can be viewed as the center of the structure. For B$_{80}$
and the NBS, the HPs and HHs act as "hollows", respectively [Figs. 1(b) and
(c)]. In light of this observation, we shall show that such a snowdrop-like
structure plays a critical role in the stability of B$_{80}$ and NBS as well
as other boron quasi-planar structures (fullerenes, sheets, nanotubes, etc),
and thus can be utilized as building blocks to construct novel boron
nanostructures. Henceforth, the boron fullerenes that are composed of
snowdrop-like structures are coined as S-fullerenes for brevity.

\begin{figure}[tbp]
\includegraphics[width=0.70\linewidth,clip]{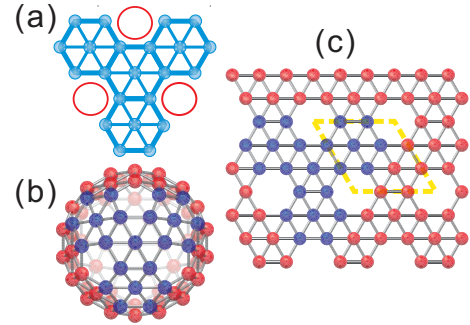}
\caption{(Color online) The snowdrop-like structures in B$_{80}$ and new
boron sheet (NBS). (a) a snowdrop-like motif, in which a filled hexagon (FH)
is surrounded by three FHs and three "hollows" (pentagons or hexagons,
denoted by circles); (b) B$_{80}$; (c) NBS, the yellow parallelogram
indicates an unit cell of the NBS. The blue atoms form the snowdrop-like
units in B$_{80}$ and NBS.}
\label{fig1}
\end{figure}

Geometrically, in an S-fullerene, every HP should be surrounded by five FHs,
and every HH is surrounded by six FHs, whereas there are three "hollows"
(HPs or HHs) around every FH. Thus we have $5N_{HP}+6N_{HH}=3N_{FH}$, where $%
N_{HP}$, $N_{HH}$, and $N_{FH}$ denote the number of HP, HH, and FH,
respectively. If we take the S-fullerene as a polyhedron, and each HP, HH or
FH as a surface, then Euler theorem, $e$ (edges) $+$ 2 $=$ $v$ (vertices) $+$
$f$ (faces), leads to $N_{HP}=12$, implying that in an S-fullerene the
number of hollow pentagons should be exactly $12$. Consequently, we have $%
N_{FH}=20+2N_{HH}$. In an S-fullerene the total number of boron atoms should
be $n =[5N_{HP}+6(N_{HH}+N_{FH})]/3+N_{FH}=80+8N_{HH}$. For B$_{80}$, since $%
N_{HH}=0$, $N_{FH}=20$, and $n=80$.

\begin{figure}[tbp]
\includegraphics[width=0.70\linewidth,clip]{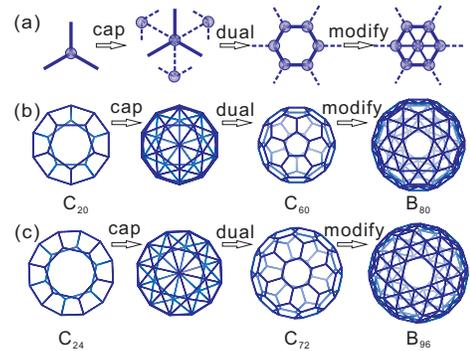}
\caption{(Color online) Illustration of leapfrog and modified leapfrog
transformations. (a) a three-connected vertex becomes a new hexagon (in
leapfrog) or a hexagon with an additional atom in center (in modified
leapfrog); (b) a C$_{20}$ converts to C$_{60}$ and B$_{80}$ by the leapfrog
and modified leapfrog transformations, respectively; (c) a C$_{24}$
transforms to C$_{72}$ and B$_{96}$.}
\label{fig2}
\end{figure}

These relations put restrictions in searching for S-fullerenes, but they are
still inadequate to find the concrete geometrical structures. Fortunately,
through intensive attempts we disclose that a modified leapfrog
transformation (MLT) is useful to generate S-fullerenes. The original
leapfrog transformation \cite{magic1} has been applied to construct the
so-called electronic closed-shell fullerenes C$_{60+6k}$
(leapfrog-fullerenes). To produce S-fullerenes, the only thing we should do
is to modify the operations shown in Fig. 2(a). In the original leapfrog
operations, every vertex transforms to an HH (containing six atoms). Here we
propose an MLT, in which every vertex converts to an FH (containing seven
atoms), suggesting that every vertex of the original polyhedron provides an
extra atom. Since these extra atoms could compose a polyhedron identical to
the original one, the MLT can be considered as the combination of the
original polyhedron and its leapfrog. For instance, B$_{80}$, the modified
leapfrog of C$_{20}$, could be viewed as a combination of C$_{20}$ and its
leapfrog C$_{60}$ structures [Fig. 2(b)]. As a result, in the final modified
leapfrog structure, the number of atoms grows four times vertices as many as
the original, i.e., $80+8k$ ($k\geq 0,k\neq 1$), while the symmetry is still
kept. The most important point is that all structures generated by the MLT
fall into the S-fullerenes. Through a reverse transformation, every
S-fullerene could be converted to the original polyhedron. Therefore, B$%
_{80+8k}$ ($k\geq 0,k\neq 1$) can be produced from the structure of C$%
_{20+2k}$ ($k\geq 0,k\neq 1$) by means of the MLT [Figs. 2 (b) and (c)]. The
isomer numbers of both C$_{60+6k}$ and B$_{80+8k}$ are equal to that of C$%
_{20+2k}$ for each $k$ ($k\geq 0,k\neq 1$), and the numbers of valence
electrons are identical for C$_{60+6k}$ and B$_{80+8k}$ with the same $k$.
In addition, the MLT may also be applied to generate other quasi-planar
structures composed of snowdrop-like units. As the leapfrog of a graphene
structure is still graphene, the MLT of graphene leads to the NBS structure;
similarly, the structures of BNTs can be generated by the MLT of CNTs.

Several structures of boron S-fullerenes as examples shown in Fig. 3 have
been generated and then studied by means of \emph{ab initio} calculations
\cite{calculation}. The isomer numbers and symmetries of B$_{80+8k}$ are
listed in Table I. For $k=6,7,8,9,10,...$, the isomer number is $%
6,6,15,17,40...$, respectively, which grows rapidly with increasing $k$. For
instance, when $k=0$, the corresponding S-fullerene is the known spherical B$%
_{80}$ \cite{B80predict}; when $k=2$, it produces B$_{96}$ [Fig.
3(a)], which bears D$_{6d}$ symmetry and has a round pillow shape
with two HHs located at the D$_{6d}$ axis; when $k=3$, it gives a
pineapple shaped B$_{104}$ with D$_{3h}$ symmetry [Fig. 3(b)]; when
$k=4$, it generates B$_{112}$ [Figs. 3(c)-(d)], which has two
isomers: one possesses D$_{2}$ symmetry with an elliptical pillow
shape, while another has T$_{d}$ symmetry and looks like a regular
tetrahedron with four smoothed vertices; when $k=5$, it yields
B$_{120}$ [Figs. 3(e)-(g)], which has three isomers with one bearing
D$_{5h}$ symmetry, and the other two having C$_{2v}$ symmetry. The
D$_{5h}$ B$_{120}$ has a cylinder capsulate shape, and can be viewed
as a short open-end BNT B$_{40}$ unit (which is rolled from NBS)
capped with two halves of B$_{80}$. This capped BNT can be elongated
by repeating the middle open-end BNT B$_{40}$ unit to
generate similar D$_{5h}$ symmetry capped BNTs B$_{80+40t}$ ($t\geq 1$) \cite%
{B180, B120}, all of which belong to a subset of B$_{80+8k}$. Generally,
using a proper S-fullerene as the cap, and the open-end BNT with different
sizes as the body, the other similar subsets of boron nanotubes could be
generated.

\begin{figure}[tbp]
\includegraphics[width=0.70\linewidth,clip]{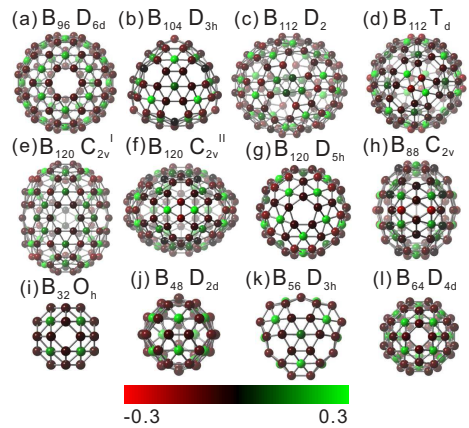}
\caption{(Color online) Schematic structures and Mulliken atomic
charge populations of several boron S-fullerenes. The unit of color
bar is electrons.} \label{fig3}
\end{figure}

\begin{table}[tbp]
\caption{The number (N$_{iso}$), symmetries, relative energies per
atom (E, in eV, relative to that of graphene and NBS, respectively),
and HOMO-LUMO gap energies ($\Delta$E, in eV) of C$_{60+6k} $ and
S-fullerenes B$_{80+8k}$
(k$\le$5) isomers (obtained with Gaussian03).}%
\begin{tabular}{ccccccccccccc}
\hline\hline
&  &  &  & \multicolumn{3}{c}{C$_{n=60+6k}$} &  & \multicolumn{4}{c}{B$%
_{n=80+8k}$} &  \\ \cline{5-7}\cline{9-13}
$k$ & N$_{iso}$ & symmetry &  & $n$ & E & $\Delta$E &  & $n$ & E & $\Delta$E
& N$_{HH}$ & N$_{FH}$ \\ \hline
0 & 1 & I$_{h}$ &  & 60 & 0.386 & 1.670 &  & 80 & 0.179 & 1.020 & 0 & 20 \\
1 & 0 & - &  & - & - & - &  & - & - & - & - & - \\
2 & 1 & D$_{6d}$ &  & 72 & 0.365 & 1.445 &  & 96 & 0.160 & 0.784 & 2 & 24 \\
3 & 1 & D$_{3h}$ &  & 78 & 0.346 & 1.460 &  & 104 & 0.151 & 0.738 &
3 & 26
\\
4 & 2 & D$_{2}$ &  & 84 & 0.339 & 1.367 &  & 112 & 0.146 & 0.682 & 4 & 28 \\
&  & T$_{d}$ &  & 84 & 0.329 & 1.632 &  & 112 & 0.141 & 0.927 & 4 & 28 \\
5 & 3 & C$_{2v}^{I}$ &  & 90 & 0.321 & 1.364 &  & 120 & 0.137 &
0.709 & 5 &
30 \\
&  & C$_{2v}^{II}$ &  & 90 & 0.322 & 1.489 &  & 120 & 0.136 & 0.828
& 5 & 30
\\
&  & D$_{5h}$ &  & 90 & 0.312 & 1.029 &  & 120 & 0.135 & 0.555 & 5 & 30 \\
\hline\hline
\end{tabular}%
\end{table}

The energies of C$_{60+6k}$ and B$_{80+8k}$ ($k\neq 1$, $k\leq 5$) relative
to graphene and NBS, respectively, are listed in Table I. Note that
different isomers with the same $k$ may have different structures and
symmetries. It can be seen that the relative energies decrease with the
increase of $k$ for both C$_{60+6k}$ and B$_{80+8k}$, which are expected to
converge to the values of graphene \cite{magic1} and NBS, respectively,
suggesting that the stabilities of both fullerenes increase with the growth
of size. Thus, B$_{80+8k}$ ($k\geq 0$, $k\neq 1$) gives rise to a large
class of stable boron fullerenes, where B$_{80}$ is just the first member of
this family, while the NBS is the infinite $k$ analogue. To check further
the stabilities of these boron fullerenes, we calculate the total energies
per atom for the caged B$_{80+8k}$ ($k\neq 1$, $k\leq 5$) with other
different structures that do not belong to the S-fullerene family, and find
that they are all less stable than those of S-fullerenes. Note that B$_{96}$
is about 0.019 eV/atom more stable than B$_{80}$ in energy, and the members
with larger $k$ have even more lower relative energies. The present class of
S-fullerenes are more stable than B$_{180}$ proposed earlier \cite{B180}.

In the preceding S-fullerenes, only are HP and HH considered as boron
"hollows". If we take the hollow quadrangle (HQ) into account, then the
S-fullerenes would be extended. For the boron fullerenes composed of HQ, HP,
HH, and FH, the total number of boron atoms satisfies $%
n=[4N_{HQ}+5N_{HP}+6(N_{HH}+N_{FH})]/3+N_{FH}$, while Euler theorem requires
$N_{HP}+2N_{HQ}=12$, and $N_{FH}=20+2(N_{HH}-N_{HQ})$. Then, we have $%
n=80+8(N_{HH}-N_{HQ})$. Consider a few special cases. (a) If $N_{HQ}=0$, it
recovers to the preceding B$_{{80+8k}}$ ($k\geq 0$, $k\neq 1$); (b) If $%
N_{HP}=0 $, we get $N_{HQ}=6$, $N_{FH}=8+2N_{HH}$, and $n=32+8N_{HH}$, which
leads to B$_{32+8k}$ ($k\geq 0$) that consist of HQs, HHs and FHs; (c) If $%
N_{HH}=0$, we have $N_{FH}=20-2N_{HQ}$, $n=80-8N_{HQ}$, and $0<N_{HQ}\leq 6$%
, which corresponds to B$_{32+8k}$ ($0\leq k<6$) that are composed of HQs,
HPs and FHs. As examples, O$_{h}$ B$_{32}$, D$_{5h}$ B$_{40}$, D$_{2d}$ B$%
_{48}$, D$_{3h}$ B$_{56}$, D$_{4d}$ B$_{64}$, and D$_{3h}$ B$_{72}$
can be obtained, as shown in Figs. 3(i)-(l); (d) If
$N_{HH}-N_{HQ}=1$, then B$_{88}$ can be formed, which has abundant
isomers because N$_{HH}$ and N$_{HQ}$ can have various combinations
to satisfy such a difference of unity. The structure of one C$_{2v}$
B$_{88}$ isomer is given in Fig. 3(h). \emph{Ab initio} calculations
show that these several isomers also follow the trend that relative
energies per boron atom decrease with the increase of $k$. Thus if
quadrangles were added to the set of "hollows", we would have a
class of fullerenes B$_{32+8k}$ ($k\geq 0$), and their isomers also
increase largely for every $k$, which can be generated by the MLT as
well, and enriches the family of boron fullerenes.

By carefully checking the structures of the S-fullerenes and NBS, we
uncover an electron counting rule (ECR) and an isolated hollow rule
(IHR) that would manifest the fundamental importance of the
snowdrop-like structure and, reasonably explain the high stability
and reveal the electronic bonding property of these boron
nanostructures. Let us look again at the structures of B$_{80}$ and
NBS. The former is composed of 12 HPs and 20 FHs, while a unit cell
of NBS has one HH and two FH [Fig. 1(c)]. As one FH can be viewed as
a group of six triangles, B$_{80}$ contains 120 triangles and an NBS
unit cell has 12 triangles. Since each boron atom has three valence
electrons, one may find that the total number of valence electrons
of B$_{80}$ is 240, just twice the number of triangles.
Interestingly, the NBS has the same electron counting feature, as
one unit cell has 8 boron atoms, contributing 24 valence electrons,
which is just also twice the number of triangles it possesses. It is
easy to substantiate that all of the S-fullerenes follow this
electron counting scheme. In the central FH of a boron snowdrop-like
unit, the atoms on six periphery vertices are shared by the
neighboring FHs, and only is the central atom monopolized, so the
net number of boron atoms that the central FH has is only four, and
the corresponding number of valence electrons is 12, just twice the
six triangles it possesses. Since all FHs in the S-fullerenes can be
viewed as the central FHs of snowdrop-like structures, being
equivalent to the above electron counting scheme, the total number
of valence electrons of S-fullerenes are twice the number of
triangles they have.

Such an agreement on electron counting property is not accidental,
which could be rationalized in a simple way \cite{ECR}. As is
well-known, the multi-center deficient-electron bonding scheme
exists widely in boron structures \cite{boron3c2e1}. Suppose that
three boron atoms on the vertices of a triangle share a three-center
two-electron bond, i.e., a triangle would consume two electrons,
then the above ECR could be properly understood in a way that, in
stable quasi-planar boron structures that are composed of FHs, HPs
and HHs, etc., the total number of the valence electrons is equal to
twice the number of triangles they contain, leading to that the
number of boron atoms is as quadruple as the number of FHs
($n=4N_{FH}$). The ECR is thus consistent with the snowdrop-like
structure, as the latter is a kind of the geometrical realization of
ECR, whilst the ECR can be considered as the electronic
interpretation of the snowdrop-like unit. On the other hand, in all
S-fullerenes, the "hollows" (HPs and HHs) are separated by FHs, as
the requirement that every FH should be the center of a
snowdrop-like structure cannot be satisfied if there exist any
adjacent "hollows". Therefore, an IHR for boron S-fullerenes can be
expected. The leapfrog procedure never generates adjacent pentagons
\cite{magic1}, and similarly, the MLT never produces adjacent
"hollows". It is interesting to note that the NBS and the related
BNTs also follow the ECR and IHR, which suggests that the above two
rules may be applicable to other boron nanostructures. While the IHR
calls for boron "hollow" be well-separated by FHs, the ECR requires
a quantitative constraint between the number of boron "hollows" and
that of FHs. Consequently, the IHR and ECR are mutually constrained,
which results in appropriate boron nanostructures.

The electronic structures of B$_{80+8k}$ are obtained by the \emph{ab initio}
calculations. Figure 3 shows the Mulliken atomic charge populations of
several S-fullerenes. The boron atoms in centers of FHs are marked in green,
showing they are positively charged, while most of others are red,
indicating they are negatively charged. Recall that all carbon atoms in C$%
_{60+6k}$ are electroneutral. With regard to electron transfers in B$%
_{80+8k} $, the central atoms of FHs act as electron donors, and the other
atoms act as electron acceptors, which are consistent with the
donor-acceptor hypothesis on NBS \cite{newboronsheet}, and also the charge
transfer in monoatomic systems \cite{monoatomic}.

\begin{figure}[tbp]
\includegraphics[width=0.75\linewidth,clip]{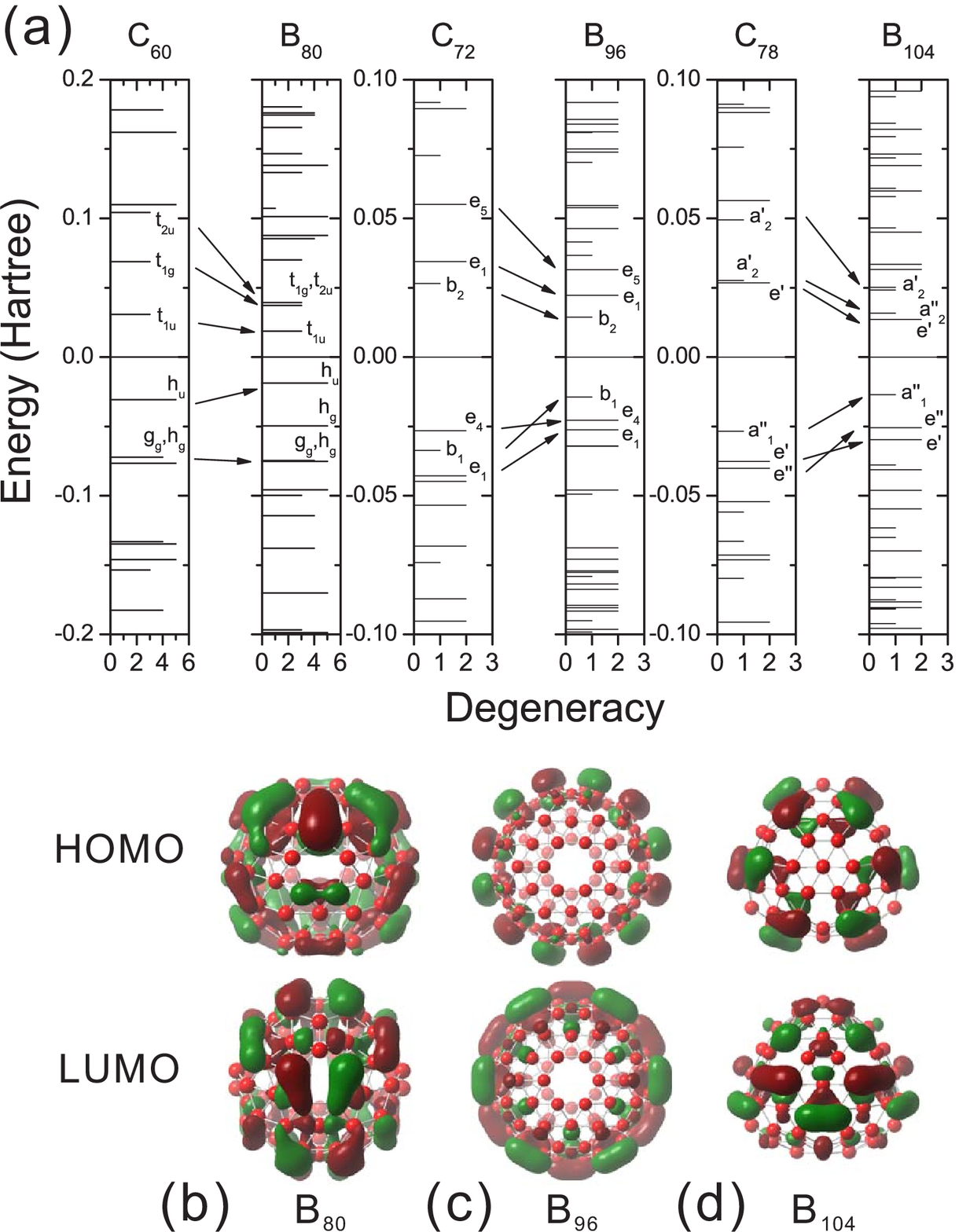}
\caption{(Color online) (a) Energy levels near HOMO-LUMO gaps of C$_{60}$, B$%
_{80}$, C$_{72}$, B$_{96}$, C$_{78}$, and B$_{104}$; The profiles of HOMO
and LUMO orbitals of (b) B$_{80}$; (c) B$_{96}$ and (d) B$_{104}$.}
\label{fig4}
\end{figure}

The gap energies between the lowest-unoccupied-molecular-orbital
(LUMO) and the highest-occupied-molecular-orbital (HOMO) of
C$_{60+6k}$ and B$_{80+8k}$ ($0\leq k\leq 5$, $k\neq 1$) are listed
in Table I. Interestingly, the isomers of both fullerenes with
$k=0(I_{h}),4(T_{d}),5(C_{2v}^{II})$ have larger gap energies than
their near neighbors. Fig. 4(a) shows the energy levels around the
HOMO-LUMO gaps of C$_{60+6k}$ and B$_{80+8k}$ ($k=0,2,3$). For the
major energy levels of B$_{80+8k}$, one may find correspondences to
their C$_{60+6k}$ counterparts with the same degeneracy and
symmetry. For instance, C$_{60}$ and B$_{80}$ both have a five-fold
degenerate $h_{u}$ HOMO and a three-fold degenerate $t_{1u}$ LUMO
\cite{Gopakumar}; C$_{72}$ and B$_{96}$, C$_{78}$ and B$_{104}$ also
exhibit similar corresponding relations as indicated in Fig. 4(a).
\emph{\ }The visualizations of the HOMO and LUMO of B$_{80+8k}$
($k=0,2,3$) are presented in Figs. 4(b)-(d). We find the main
profiles of corresponding MOs (C$_{60+6k}$ MOs are not shown here)
exhibit striking similarity between C$_{60}$ and B$_{80}$, C$_{72}$
and B$_{96}$, C$_{78}$ and B$_{104}$, and so forth. We believe that
such an observation from the first several members of B$_{80+8k}$
family would be kept for the members with even larger $k$. Since the
frontier MOs dominate the primary chemical property of a molecule,
B$_{80+8k}$ may have chemical properties similar to the corresponding C$%
_{60+6k}$. Besides, as the carbon fullerenes can be stabilized by the
exohedral chemical deriving and metallic endohedral method \cite{Xie, Wang,
Yan}, the stability of boron fullerenes might also be improved in the same
manner.

In summary, the general constructing schemes were proposed to construct a
large family of stable boron fullerenes, whose remarkable stabilities were
confirmed by intensive first-principles simulations. The empirical ECR and
IHR were suggested to readily explain the geometrical stability and reveal
the electronic bonding of the predicted boron fullerenes, which can also be
shown applicable to boron quasi-planar structures such as nanotubes and
sheets, and be useful for seeking for other stable boron nanostructures. The
present predictions are awaited to synthesize experimentally, which may
bring about a new dimension of boron chemistry.

%The authors are grateful to X. Chen, S. S. Gong, Y. Wang, Z. C.
%Wang, Z. Xu, G. Q. Zhong for useful discussion and help.

All of the calculations are completed on the supercomputer NOVASCALE 6800 in
Virtual Laboratory of Computational Chemistry, Supercomputing Center of
Chinese Academy of Sciences. This work is supported in part by the National
Science Foundation of China
%(Grant Nos. 90403036, 20490210), the National Science Fund for
%Distinguished Young Scholars of China
(Grant No. 10625419). % the MOST of China (Grant No. 2006CB601102),
%and the Chinese Academy of Sciences.


\begin{thebibliography}{99}
\bibitem{calculation} All of calculations were performed using Gaussian 03
\cite{Gaussian} with PBEPBE density functional \cite{PBE} and all-electron
6-31G(d, p) basis set \cite{basis} employed. The geometrical structures are
optimized until the total energy and force achieve the tolerance of about 1$%
\times$10$^{-8}$ Hartree and 4.5$\times$10$^{-4}$ Hartree/Bohr,
respectively. The relative energies and HOMO-LUMO gaps were rechecked by
using Quantum-ESPRESSO package \cite{PWSCF} with ultrasoft Vanderbilt
pseudopotentials \cite{USPP} and generalized gradient corrected exchange and
correlation functional PBE \cite{PBE} employed. The results agree quite well.
\end{thebibliography}
\end{document}